\documentclass[twocolumn]{aastex62}
\usepackage{longtable}
\usepackage{comment}
\usepackage{amsmath, bm, nccmath}

\newcommand{\feh}{\ensuremath{{\rm [Fe/H]}}}
\newcommand{\teff}{\ensuremath{T_{\rm eff}}}
\newcommand{\teq}{\ensuremath{T_{\rm eq}}}
\newcommand{\logg}{\ensuremath{\log{g}}}
\newcommand{\zaspe}{\texttt{ZASPE}}
\newcommand{\ceres}{\texttt{CERES}}
\newcommand{\tess}{\textit{TESS}}
\newcommand{\vsini}{\ensuremath{v \sin{i}}}
\newcommand{\kms}{\ensuremath{{\rm km\,s^{-1}}}}
\newcommand{\mjup}{\ensuremath{{\rm M_{J}}}}
\newcommand{\mearth}{\ensuremath{{\rm M}_{\oplus}}}
\newcommand{\mpl}{\ensuremath{{\rm M_p}}}
\newcommand{\rjup}{\ensuremath{{\rm R_J}}}
\newcommand{\rpl}{\ensuremath{{\rm R_P}}}
\newcommand{\rstar}{\ensuremath{{\rm R}_{\star}}}
\newcommand{\mstar}{\ensuremath{{\rm M}_{\star}}}

\newcommand{\rsun}{\ensuremath{{\rm R}_{\odot}}}
\newcommand{\msun}{\ensuremath{{\rm M}_{\odot}}}
\newcommand{\lsun}{\ensuremath{{\rm L}_{\odot}}}

\newcommand{\stlum}{\ensuremath{2.33 \pm 0.11 }}
\newcommand{\stage}{\ensuremath{2.92_{-0.73}^{+0.80} }}
\newcommand{\av}{\ensuremath{0.226_{-0.058}^{+0.056} }}
\newcommand{\stteff}{\ensuremath{6295 \pm 77 }}
\newcommand{\stfeh}{\ensuremath{0.00 \pm 0.05 }}
\newcommand{\stlogg}{\ensuremath{4.291 \pm 0.025 }}
\newcommand{\stvsini}{\ensuremath{7.80 \pm 0.19 }}
\newcommand{\stmstan}{\ensuremath{1.170\pm0.06}}
\newcommand{\strstan}{\ensuremath{1.282\pm0.03}}

\newcommand{\per}{\ensuremath{11.23660 \pm 0.00011}}
\newcommand{\bb}{\ensuremath{0.723_{-0.024}^{+0.018}}}
\newcommand{\rr}{\ensuremath{0.0942_{-0.0012}^{+0.0010}}}
\newcommand{\omegapost}{\ensuremath{1.230_{-0.063}^{+0.063}}}
\newcommand{\ecc}{\ensuremath{0.435\pm 0.024 }}
\newcommand{\logm}{\ensuremath{5.973_{-0.056}^{+0.054}}}
\newcommand{\mptess}{\ensuremath{1.236_{-0.067}^{+0.069}}}

\newcommand{\rptessrjup}{\ensuremath{1.170\pm 0.03 }}
\newcommand{\stm}{\ensuremath{1.181\pm 0.058 }}
\newcommand{\str}{\ensuremath{1.28_{-0.03}^{+0.03}}}
\newcommand{\aone}{\ensuremath{1.58\pm 0.19 }}
\newcommand{\azero}{\ensuremath{-6.4_{-8.2}^{+8.1}}}
\newcommand{\bone}{\ensuremath{0.0204_{-0.0040}^{+0.0039}}}
\newcommand{\bzero}{\ensuremath{-0.00197_{-0.00032}^{+0.00033}}}
\newcommand{\etaferos}{\ensuremath{13.5_{-3.1}^{+3.3}}}
\newcommand{\etacoralie}{\ensuremath{45.1_{-11.4}^{+12.4}}}
\newcommand{\etachiron}{\ensuremath{21.7_{-7.6}^{+8.0}}}
\newcommand{\etaminerva}{\ensuremath{33.0_{-3.8}^{+4.5}}}
\newcommand{\etassmko}{\ensuremath{0.00114_{-0.00010}^{+0.00011}}}
\newcommand{\gammaferos}{\ensuremath{-42.8_{-8.6}^{+8.8}}}
\newcommand{\gammacoralie}{\ensuremath{-26.9_{-15.4}^{+14.7}}}
\newcommand{\gammachiron}{\ensuremath{-20.9_{-10.6}^{+10.7}}}
\newcommand{\gammanres}{\ensuremath{-28.5_{-11.4}^{+11.4}}}
\newcommand{\gammaminervathree}{\ensuremath{4.2_{-10.2}^{+10.1}}}
\newcommand{\gammaminervafour}{\ensuremath{-32.2_{-10.5}^{+10.6}}}
\newcommand{\gammaminervafive}{\ensuremath{20.9_{-12.9}^{+12.7}}}
\newcommand{\uone}{\ensuremath{0.50_{-0.27}^{+0.19}}}
\newcommand{\utwo}{\ensuremath{-0.06_{-0.23}^{+0.33}}}
\newcommand{\rone}{\ensuremath{0.0942_{-0.0012}^{+0.0010}}}
\newcommand{\rtwo}{\ensuremath{0.723_{-0.024}^{+0.018}}}
\newcommand{\strho}{\ensuremath{0.80_{-0.06}^{+0.06}}}
\newcommand{\tzero}{\ensuremath{2458547.47448_{-0.00029}^{+0.00028}}}
\newcommand{\sma}{\ensuremath{0.1038_{-0.0017}^{+0.0017}}}
\newcommand{\inc}{\ensuremath{87.63_{-0.1}^{+0.11}}}

\newcommand{\teqv}{\ensuremath{1252\pm 21 }}

\newcommand{\plname}{TOI-677~b}
\newcommand{\stname}{TOI-677}
\newcommand{\stnameTIC}{TIC~280206394}

\newcommand{\rhostar}{\ensuremath{\rho_*}}

\graphicspath{{./}{figures/}}

\received{}
\revised{}
\accepted{}
\submitjournal{AAS}

\shorttitle{An Eccentric Warm Jupiter}
\shortauthors{Jord\'an et al.}

\begin{document}

\title{\plname: A Warm Jupiter (P=11.2d) on an eccentric orbit transiting a late F-type star}

\correspondingauthor{Andr\'es Jord\'an}
\email{andres.jordan@uai.cl}

\author[0000-0002-5389-3944]{Andr\'es Jord\'an}
\affiliation{Facultad de Ingeniería y Ciencias, Universidad Adolfo Ib\'a\~nez, Av.\ Diagonal las Torres 2640, Pe\~nalol\'en, Santiago, Chile}
\affiliation{Millennium Institute for Astrophysics, Chile}

\author[0000-0002-9158-7315]{Rafael Brahm}
\affiliation{Center of Astro-Engineering UC, Pontificia Universidad Cat\'olica de Chile, Av. Vicu\~{n}a Mackenna 4860, 7820436 Macul, Santiago, Chile}
\affiliation{Instituto de Astrof\'isica, Pontificia Universidad Cat\'olica de Chile, Av.\ Vicu\~na Mackenna 4860, Macul, Santiago, Chile}
\affiliation{Millennium Institute for Astrophysics, Chile}

\author[0000-0001-9513-1449]{N\'estor Espinoza}
\affiliation{Space Telescope Science Institute, 3700 San Martin Drive, Baltimore, MD 21218, USA}

\author{Thomas Henning}
\affiliation{Max-Planck-Institut f\"ur Astronomie, K\"onigstuhl 17, Heidelberg 69117, Germany }

\author{Mat\'ias I.\ Jones}
\affiliation{European Southern Observatory, Alonso de C\'ordova 3107, Vitacura, Casilla 19001, Santiago, Chile}

\author{Diana Kossakowski}
\author{Paula Sarkis}
\author{Trifon Trifonov}
\affiliation{Max-Planck-Institut f\"ur Astronomie, K\"onigstuhl 17, Heidelberg 69117, Germany }

\author{Felipe Rojas}
\author{Pascal Torres}
\affiliation{Instituto de Astrof\'isica, Pontificia Universidad Cat\'olica de Chile, Av.\ Vicu\~na Mackenna 4860, Macul, Santiago, Chile}
\affiliation{Millennium Institute for Astrophysics, Chile}

\author{Holger Drass}
\affiliation{Center of Astro-Engineering UC, Pontificia Universidad Cat\'olica de Chile, Av. Vicu\~{n}a Mackenna 4860, 7820436 Macul, Santiago, Chile}

\author{Sangeetha Nandakumar}
\author{Mauro Barbieri}
\affiliation{INCT, Universidad de Atacama, calle Copayapu 485, Copiap\'o, Atacama, Chile}

\author{Allen Davis}
\author{Songhu Wang}
\affiliation{Department of Astronomy, Yale University, New Haven, CT 06511, USA}

\author[0000-0001-6023-1335]{Daniel Bayliss}%
\affiliation{Department of Physics, University of Warwick, Coventry CV4 7AL, UK}

% POC, SPOC, TSO authors
% TSO
\author{Luke Bouma}
\affiliation{Department of Astrophysical Sciences, Princeton University, 4 Ivy Lane, Princeton, NJ 08540, USA}
\author{Diana Dragomir}
\affiliation{Department of Physics and Astronomy, University of New Mexico, 1919 Lomas Blvd NE, Albuquerque, NM 87131, USA}
\author{Jason D.\ Eastman}
\affiliation{Harvard-Smithsonian Center for Astrophysics, 60 Garden St, Cambridge, MA 02138, USA}
% POC
\author[0000-0002-6939-9211]{Tansu Daylan}
\affiliation{Department of Physics and Kavli Institute for Astrophysics and Space Research, Massachusetts Institute of Technology, Cambridge, MA 02139, USA}
\affiliation{Kavli fellow}
\author[0000-0002-5169-9427]{Natalia Guerrero}
\affiliation{Department of Physics and Kavli Institute for Astrophysics and Space Research, Massachusetts Institute of Technology, Cambridge, MA 02139, USA}
\author[0000-0001-7139-2724]{Thomas Barclay}
\affiliation{NASA Goddard Space Flight Center, 8800 Greenbelt Rd, Greenbelt, MD 20771, USA}
\affiliation{University of Maryland, Baltimore County, 1000 Hilltop Cir, Baltimore, MD 21250, USA}

%SPOC
\author{Eric B.\ Ting}
\author{Christopher E.\ Henze}
\affiliation{NASA Ames Research Center, Moffett Field, CA, 94035, USA}

% TESS architects
\author{George Ricker} %TESS architect
\affiliation{Department of Physics and Kavli Institute for Astrophysics and Space Research, Massachusetts Institute of Technology, Cambridge, MA 02139, USA}
\author{Roland Vanderspek}  %TESS architect
\affiliation{Department of Physics and Kavli Institute for Astrophysics and Space Research, Massachusetts Institute of Technology, Cambridge, MA 02139, USA}
\author{David W.\ Latham}  %TESS architect
\affiliation{Harvard-Smithsonian Center for Astrophysics, 60 Garden St, Cambridge, MA 02138, USA}
\author[0000-0002-6892-6948]{Sara Seager}  %TESS architect
\affiliation{Department of Physics and Kavli Institute for Astrophysics and Space Research, Massachusetts Institute of Technology, Cambridge, MA 02139, USA}
\affiliation{Department of Earth, Atmospheric and Planetary Sciences, Massachusetts Institute of Technology, Cambridge, MA 02139,USA}
\affiliation{Department of Aeronautics and Astronautics, MIT, 77 Massachusetts Avenue, Cambridge, MA 02139, USA}
\author{Joshua Winn} %TESS architect
\affiliation{Department of Astrophysical Sciences, Princeton University, 4 Ivy Lane, Princeton, NJ 08540, USA}
\author[0000-0002-4715-9460]{Jon M.\ Jenkins} %TESS architect
\affiliation{NASA Ames Research Center, Moffett Field, CA, 94035, USA}

% Minerva Team
\author[0000-0001-9957-9304]{Robert A.\ Wittenmyer}
\affiliation{University  of Southern Queensland, Centre for Astrophysics, West Street, Toowoomba, QLD 4350 Australia}
\author[0000-0003-2649-2288]{Brendan P.\ Bowler}
\affiliation{Department  of Astronomy, The University of Texas at Austin, TX 78712, USA}
\author{Ian  Crossfield}
\affiliation{Department  of Physics, Massachusetts Institute of Technology, Cambridge, MA, USA}
\author[0000-0002-1160-7970]{Jonathan Horner}
\affiliation{University  of Southern Queensland, Centre for Astrophysics, West Street, Toowoomba, QLD 4350 Australia}
\author[0000-0002-7084-0529]{Stephen R.\ Kane}
\affiliation{Department of Earth and Planetary Sciences, University of California, Riverside, CA 92521, USA}
\author[0000-0003-0497-2651]{John F.\ Kielkopf}
\affiliation{Department of Physics and Astronomy, University of Louisville, Louisville, KY 40292, USA}
\author{Timothy D.\ Morton}
\affiliation{Department  of Astronomy, University of Florida, 211 Bryant Space Science Center, Gainesville, FL, 32611, USA}
\author{Peter  Plavchan}
\affiliation{Department of Physics and Astronomy, George  Mason University, 4400 University Drive MS 3F3, Fairfax, VA 22030, USA}
\author{C.G.\  Tinney}
\affiliation{Exoplanetary  Science at UNSW, School of Physics, UNSW Sydney, NSW 2052, Australia}

\author{Brett Addison}
\author[0000-0002-7830-6822]{Matthew  W.\ Mengel}
\author{Jack  Okumura}
\affiliation{University of Southern Queensland, Centre for Astrophysics, West Street, Toowoomba, QLD 4350 Australia}

\author{Sahar Shahaf}
\author{Tsevi Mazeh}
\affiliation{School of Physics and Astronomy, Tel Aviv University, Tel Aviv 69978, Israel}

\author[0000-0003-2935-7196]{Markus Rabus}
\affiliation{Las Cumbres Observatory Global Telescope Network, Santa Barbara, CA 93117, USA}
\affiliation{Department of Physics, University of California, Santa Barbara, CA 93106-9530, USA}

\author{Avi Shporer}
\affiliation{Department of Physics and Kavli Institute for Astrophysics and Space Research, Massachusetts Institute of Technology, Cambridge, MA 02139, USA}

\author{Carl Ziegler}
\affil{Dunlap Institute for Astronomy and Astrophysics, University of Toronto, 50 St. George Street, Toronto, Ontario M5S 3H4, Canada}

\author[0000-0003-3654-1602]{Andrew W.\ Mann}
\affiliation{Department of Physics and Astronomy, The University of North Carolina at Chapel Hill, Chapel Hill, NC 27599-3255, USA}

\author{Rhodes Hart}
\affiliation{University
 of Southern Queensland, Centre for Astrophysics, West Street, Toowoomba, QLD 4350 Australia}

\begin{abstract}
We report the discovery of \plname, first identified as a candidate in light curves obtained within Sectors 9 and 10 of the Transiting Exoplanet Survey Satellite (\tess) mission and confirmed with radial velocities. \plname\ has a mass of \mpl = \mptess\ \mjup, a radius of \rpl = \rptessrjup\ \rjup, and orbits its bright host star ($V = 9.8$ mag) with an orbital period of \per\ d, on an eccentric orbit with $e=\ecc$. The host star has a mass of \mstar = \stm\ \msun, a radius of \rstar = \str\ \rsun, an age of \stage\ Gyr and solar metallicity, properties consistent with a main sequence late F star with $\teff=\stteff$ K.
We find evidence in the radial velocity measurements of a secondary long term signal which could be due to an outer companion. The \plname\ system is a well suited target for Rossiter-Mclaughlin observations that can constrain migration mechanisms of close-in giant planets.
\end{abstract}

\keywords{planetary systems -- stars: individual: \stname\ -- planets and satellites: gaseous planets -- planets and satellites: detection}

\section{Introduction}
\label{sec:int}

In the past two decades the population of known transiting exoplanets has grown at an accelerating pace. While the \textit{Kepler} satellite \citep{borucki:2010} dominates the overall number of discoveries, the particular class of close-in gas giants around nearby stars were until recently most efficiently discovered by wide-field photometric series \citep[e.g.][]{bakos:2004,pollacco:2006,pepper:2007,bakos:2013,talens:2017}. Due to the biases inherent to ground based observatories, most of the discoveries of these surveys have periods $P\lesssim 10$d. Systems of stations around the globe such as the HATSouth survey can in principle improve the efficiency of discovery for longer periods, but the number of systems with $P>10$d uncovered by wide-field ground-based surveys is small, with the current record holder being HATS-17b with $P\approx 16$d \citep{brahm:2016:hs17}.

The population of close-orbiting gas giants has opened a number of questions about their physical structural and dynamical evolution which are still topics of active research \citep{dawson:2018}. In particular, the nature of the migration history and the detailed mechanism of radius inflation for hot Jupiters need further elucidation. In order to make further progress on those fronts the population of \textit{warm} giants, loosely defined as systems with periods $P\gtrsim 10$d, is of importance. They are close enough to the star that they are likely to have undergone significant migration, but not as close that tidal effects can erase the potential imprints of that migration \citep{albrecht:2012,dawson:2014,gongjie:2016}. In the same vein, they are far enough from their parent star that their radii have not been inflated by the mechanism that acts to bloat the radii of hotter giants \citep{kovacs:2010, demory:2011,miller:2011}. But while it is clear that these systems are very interesting, the population of known warm giants around nearby stars (allowing the most detailed characterization) is still very small. The launch of the TESS mission \citep{ricker:2015} is changing that. By scanning nearby stars around the whole sky the expectation is that hundreds of giant planets with $P\gtrsim 10$ d will be uncovered \citep{sullivan:2015,barclay:2018}.

In this work we present the discovery originating from a TESS light curve of an eccentric warm giant planet with a period of $P=\per$ days orbiting a bright late F star. This is part of a systematic effort to characterize warm giants in the southern hemisphere uncovered with TESS which has contributed to the discovery and mass measurement of three warm giants already \citep{brahm:2019, huber:2019, rodriguez:2019}. The paper is structured as follows. In \S~\ref{sec:obs} we describe the observational material which gets used to perform a global modeling of the system as described in \S~\ref{sec:ana}. The results are then discussed in \S~\ref{sec:dis}.

\section{Observations} \label{sec:obs}

\subsection{TESS}
\label{ssec:tess}

Between 2019 March 01 and 2019 April 22, the \tess\ mission
observed \stname\ (\stnameTIC, 2MASS J09362869-5027478, TYC 8176-02431-1, WISE J093628.65-502747.3)
during the monitoring of Sectors 9 and 10, using camera 3 and CCDs 1 and 2, respectively.
The  TESS Science Processing Operations Center \citep[SPOC; for an overview of the processing it carries out see][]{jenkinsSPOC:2016} Transiting Planet Search module detected the planetary signature in the Sector 9 processing run and in the Sectors 1-13 multi-sector search and triggered the Data Validation module \citep{twicken:2018,Li:DVmodelFit:2019} to analyze the transit-like feature in the Sector 9 and combined Sectors 9 and 10 light curves. All diagnostics tests performed as part of the data validation report, including the odd/even depth test, the signal to noise ratio, the impact parameter, the statistical bootstrap probability, the ghost diagnostic, and the difference image centroid offset from the TIC position and from the out-of-transit centroid, strongly favored the planetary hypothesis and resulted in the promotion of \stname\ to the list of targets of interest.

The properties of \stname\ as obtained from literature sources and derived in this work are detailed in Table~\ref{tab:stprops}.
The target was observed in short (2 min) cadence, and we downloaded the PDC (Pre-search Data Conditioning) SAP light curves from the Mikulski Archives for Space Telescopes. The PDC SAP light curves have systematic trends removed by the use of co-trending basis vectors \citep{smithPDC:2012,stumpePDC:2014}, and are produced by the TESS SPOC at NASA Ames Research Center.
We masked the regions of high scattered light as indicated in the data release notes for each of the sectors, augmenting the masked windows in a few cases where it was evident that there were some remaining trends that were insufficiently masked\footnote{In detail, in the first and second orbits of sector 10 we excluded up to cadence numbers 247000 and 257300, respectively, instead of the values 246576 and 256215 indicated in the data release notes for sector 10.}. We did not mask datapoints with data quality flags, as we noticed that all of the second transit had been masked with a flag value of 2048 (stray light from Earth or Moon in camera field of view), but inspection of the masked portions revealed no anomalous signs on the light curve.

\begin{figure*}
\plotone{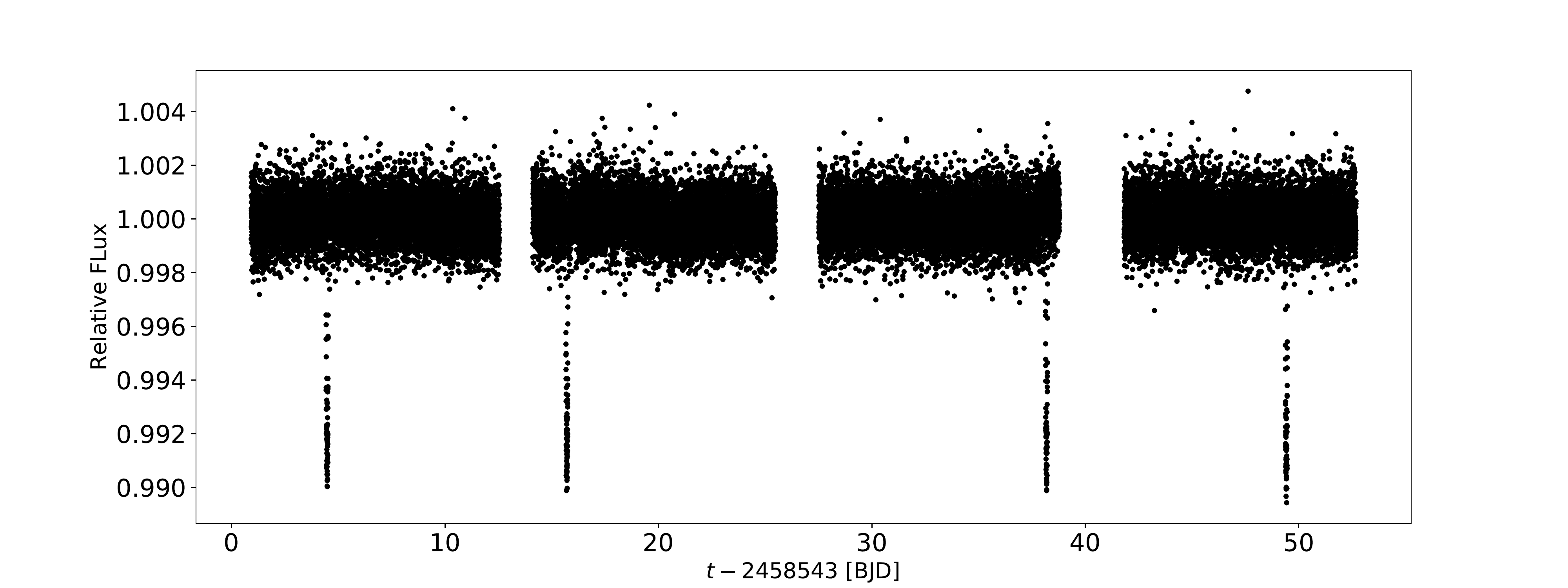}
\caption{The 2 min cadence TESS light curve. These are the PDC SAP measurements produced by the SPOC pipeline. Four transits are clearly seen in the TESS photometry, with the middle transit falling within a gap.
}
\label{fig:TESS}
\end{figure*}

\begin{figure}
\plotone{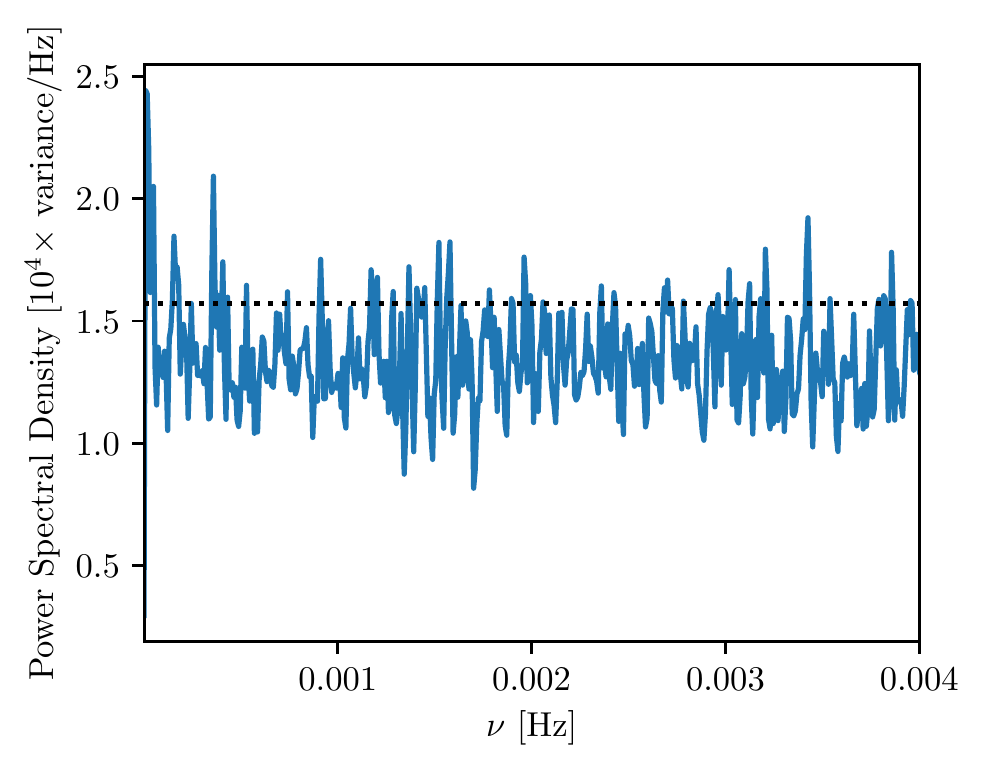}
\caption{Power spectral density of the out-of-transit TESS light curve. The dashed line marks the expected value of the power spectral density, estimated as $\langle \sigma_i^2\rangle / (\nu_u-\nu_l)$, where $\langle \sigma_i^2\rangle$ is the average measurement variance of the TESS photometry, $\nu_l=0$ and $\nu_u\approx 0.004$ Hz is the Nyquist frequency for the short cadence sampling.
}
\label{fig:PSD}
\end{figure}

The TESS light curve is shown in Figure~\ref{fig:TESS}, where four transits are clearly seen. The out-of-transit light curve is remarkably flat. We estimated the power spectral density of the out-of-transit light curve of \stname\ using the method of \citet{welch:1967} as implemented in the \textsf{scipy.signal} Python module and found it to be featureless and at precisely the level expected given the reported photometric uncertainties of the magnitude measurements (see Figure~\ref{fig:PSD}). We conclude from this exercise that there is no need for any deterministic or stochastic component beyond the white noise implied by the photometric uncertainties in the modeling of the out-of-transit light curve. Because of this we only fit for regions of $\approx$ 1d around each transit, removing the median value calculated in the out-of-transit portion for each transit. The TESS data used for the analysis is presented in Table~\ref{tab:phot}

\subsection{Spectroscopy}

We followed up \stname\ with several spectrographs in order to confirm the TESS transiting planet candidate and to measure its mass. In what follows we describe the observations obtained by each spectrograph we used. The derived radial velocities, and bisector span measurements when available, are reported in Table~\ref{tab:rvs}.

\subsubsection{FEROS}
\stname\ was monitored with the FEROS spectrograph \citep[R$\approx$48000,][]{kaufer:99} mounted at the MPG 2.2m telescope at La Silla Observatory between May and July of 2019, where 26 spectra were obtained. Observations were performed in simultaneous calibration mode, with the secondary fibre observing a thorium-argon (ThAr) lamp to trace the instrumental variations produced by changes in the environment during the science exposures. The adopted exposure times were of 300 s and 400 s, which translated into a signal-to-noise ratio ranging between 40 and 150 per resolution element. The FEROS data were processed with the \ceres\ pipeline \citep{brahm:2017:ceres}, which delivers the radial velocities corrected by the instrumental drift variations and the by the Earth's motion. These radial velocities were obtained with the cross-correlation technique, where a G2-type binary mask was used as template.
From this cross-correlation peak \ceres\ also computes the bisector span measurements, and delivers a rough estimate of the stellar parameters by comparing the continuum normalized spectrum with a grid of synthetic ones.

\subsubsection{Coralie}

We monitored \stname\ with the Coralie spectrograph \citep[R$\approx$60000,][]{mayor:2003} mounted on the Swiss-Euler 1.2m telescope in six different epochs. These observations were also performed with the simultaneous calibration technique, but in this case the secondary fibre is illuminated by a Fabry-Perot etalon. We adopted an exposure time of 300 s, which produced spectra having a typical signal-to-noise ratio of 30 per resolution element. Coralie data were also processed with the \ceres\ pipeline for obtaining the radial velocities.

\subsubsection{CHIRON}

We collected a total of 11 spectra of \stname\ using the CHIRON high-resolution spectrograph \citep{tokovinin:2013}, between May 17 and June 19, 2019. The exposure time was between 750--1200 s, leading to a signal-to-noise ratio (SNR) per pixel between $\approx$
20\,-\,35. CHIRON is mounted on the SMARTS 1.5\,m telescope at the Cerro Tololo Inter-American observatory in Chile, and is fed by an octagonal multi-mode optical fibre. For these observations we used the image slicer, which delivers relatively high throughput and high spectral resolution (R $\approx$ 80,000). The radial velocities were computed from the cross-correlation function between the individual spectra and a high-resolution template of the star, which is built by stacking all individual observations of this star. Since CHIRON is not equipped with a simultaneous calibration, we observed the spectrum of a Th-Ar lamp before the science observations, to correct for the instrumental drift.
Using this method we have measured a long-term RV stability of $<$ 10 m\,s$^{-1}$ on bright targets (t$_{exp} <$ 60 s) and $<$ 15 m\,s$^{-1}$ for fainter objects (t$_{exp} <$ 1800 s). For more details of the method see \citet{wang:2019} and  \citet{jones:2019}.

\subsubsection{NRES}

NRES \citep{nres:2018} is a global array of echelle spectrographs mounted on 1-meter telescopes, with a resolving power of $\approx$ 53,000. \stname\ was observed at 12 epochs with the NRES node located at the Cerro Tololo Inter-American Observatory. At each observing epoch, three consecutive 1200 s exposures were obtained, with individual signal-to-noise ratio $\gtrsim$40. The velocity of each exposure was derived via cross-correlation with a PHOENIX template \citep{husser:2013} with \teff=5800 K, \logg=3.5, \feh=-0.5 and \vsini=7 km/s. Systematic drifts were corrected per order \citep[e.g.,][]{engel:2017} and the radial velocity of each epoch was then taken as the mean of the three exposures.

\subsubsection{Minerva-Australis}

We obtained 17 observations on nine separate nights with the MINERVA-Australis telescope array \citep{addison:2019} at Mount Kent Observatory in Queensland, Australia. All of the telescopes in the MINERVA-Australis array simultaneously feed a single Kiwispec R4-100 high-resolution (R$\approx$80,000) spectrograph with wavelength coverage from 500 to 630 nm over 26 echelle orders. We derived radial velocities for each telescope using the least-squares analysis of \citet{anglada:2012}  and corrected for spectrograph drifts with simultaneous Thorium-Argon arc lamp observations. \stname\ was observed with telescopes 3, 4 and 5 of the array, the derived radial velocities are reported under the instrument labels Minerva\_T3, Minerva\_T4 and Miverva\_T5 in Table~\ref{tab:rvs}.

\subsection{Ground-based photometry}
\label{ssec:ground}

\subsubsection{Shared Skies Telescope at Mt.\ Kent Observatory (SSMKO)}

\begin{figure}
\plotone{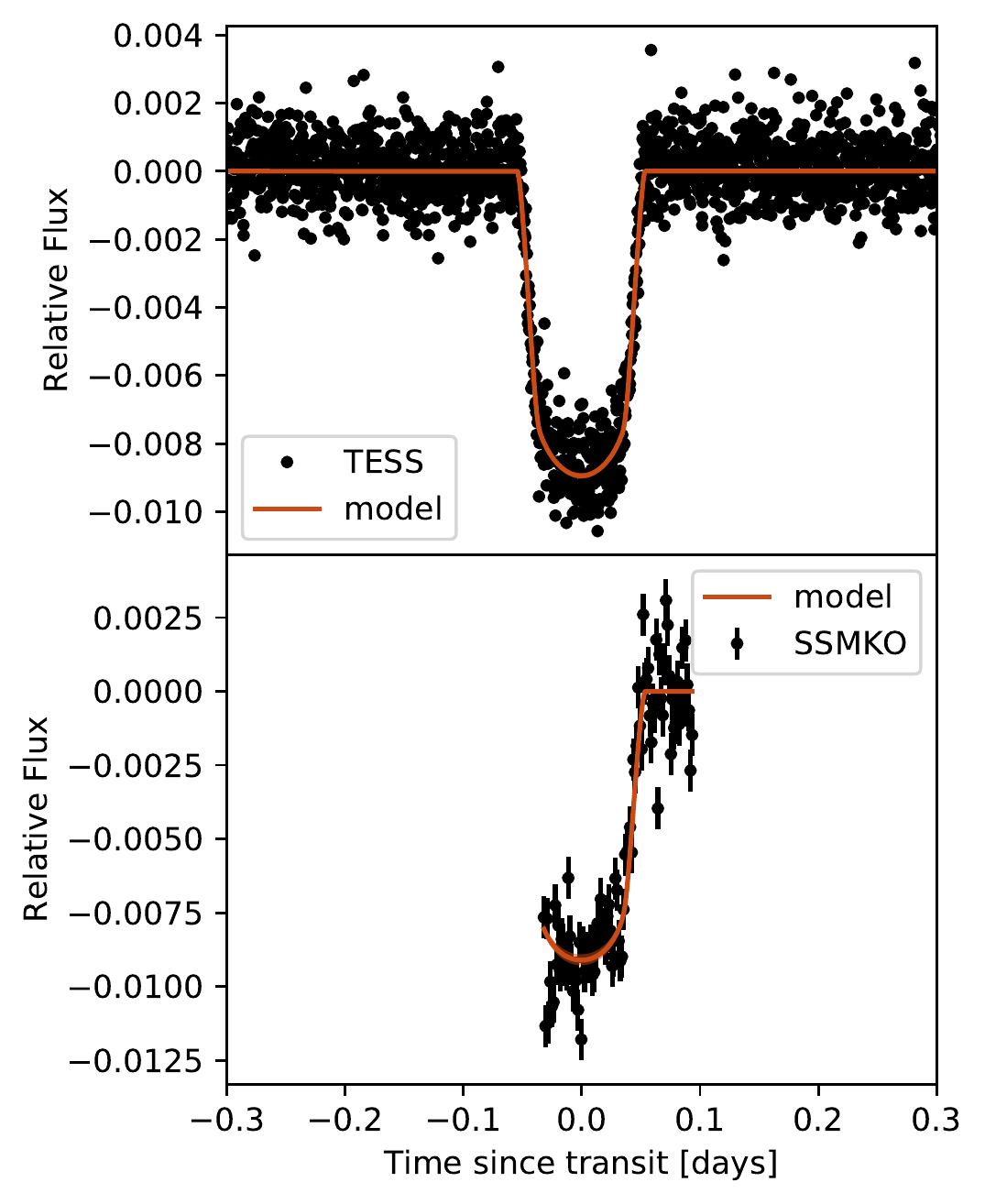}
\caption{Photometric data with the trend subtracted as a function of phase for the two photometric instruments used in this work (TESS, SSMKO). The orange line shows the posterior transit model.
}
\label{fig:lcs_phased}
\end{figure}

\stname\ was observed on the night of UTC 2019-05-09 with the University of Louisville's Shared Skies MKO-CDK700 (SSMKO) telescope at Mt.\ Kent Observatory of the University of Southern Queensland, Australia.  The telescope is a 0.7-meter corrected Dall-Kirkham with a Nasmyth focus manufactured by Planewave. Images with an exposure time of 64 s were taken through a Sloan i' filter using an Apogee U16 CCD camera with a Kodak KAF-16801E sensor. A sequence of 92 images were acquired over 180 minutes. The light curve, which is shown in  Figure~\ref{fig:lcs_phased} displays a clear egress. No significant activity or modulation other than the transit itself was apparent in the light curve, which shows residuals of 0.85 ppt at the observational cadence. The SSMKO data used for the analysis is presented in Table~\ref{tab:phot}

\subsection{Gaia DR2}

Observations of \stname\ by Gaia were reported in DR2 \citep{gaia, gaia:dr2}. From GAIA DR2, \stname\ has a parallax of $7.02 \pm 0.03$ mas,
an effective temperature of $\teff = 5895_{-200}^{+105}$ K and a radius
of $\rstar = 1.37_{-0.05}^{+0.1} \,\, \rsun$. The
parallax obtained from GAIA was used to determine the stellar physical parameters of \stname\ as described in Section~\ref{ssec:stellarpars}.  In our analysis we corrected the GAIA DR2 parallax for the systematic offset of -82 $\mu$as reported in \citet{stassun:2018}.

%ra = 144.11935935812366
%dec = -50.46309333786793
%GAIA source id: 5313031504751276032

\subsection{High spatial resolution imaging}

\begin{figure}
\plotone{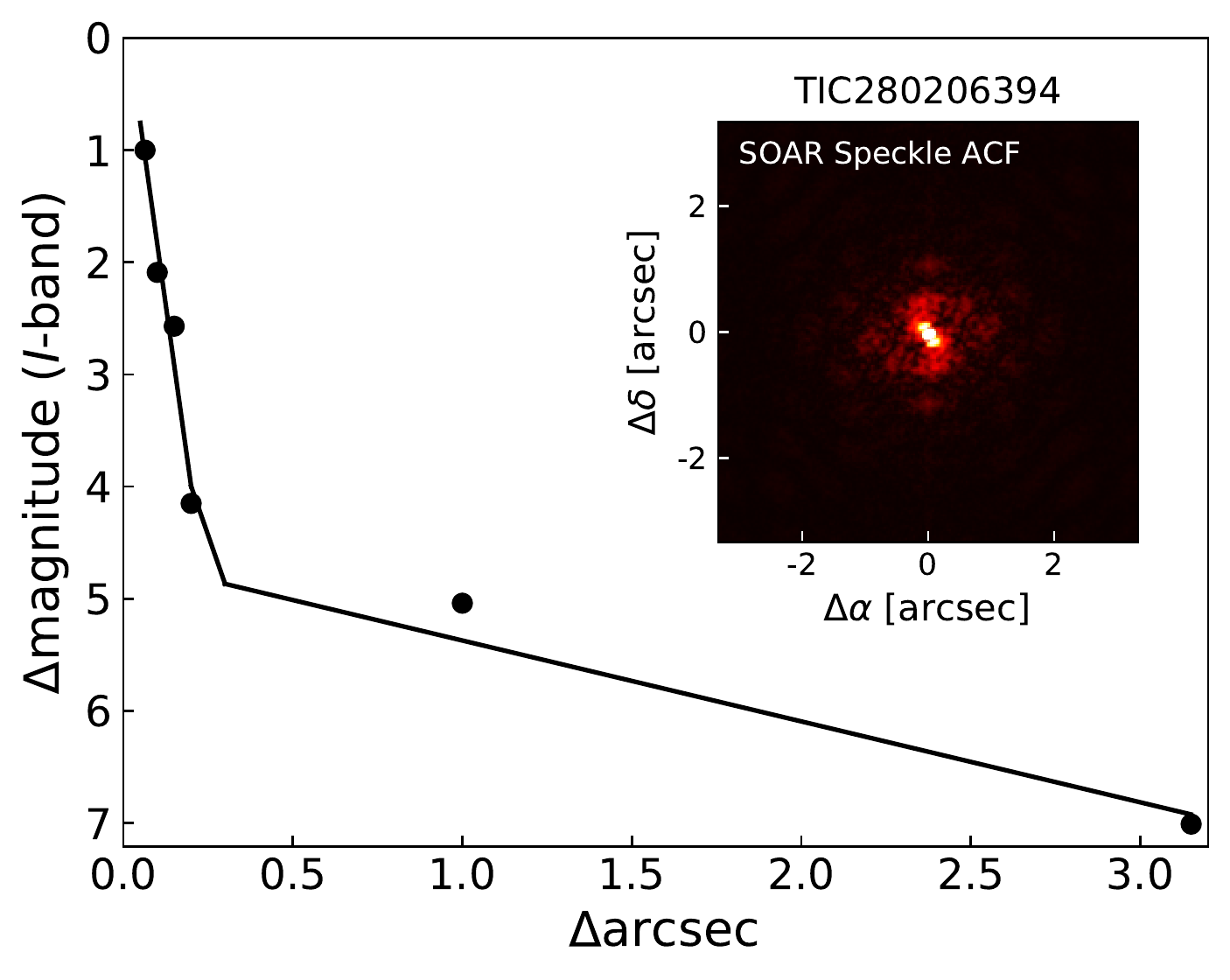}
\caption{The $I$-band auto-correlation function from Speckle using SOAR. The 5-$\sigma$ contrast curve for \stname\ is shown by the black points. The black solid line is the linear fit to the data for separations $<0\farcs2$ and $>0\farcs2$. The auto-correlation function is shown within the contrast curve plot.
}
\label{fig:soar}
\end{figure}

The relatively large angle subtended by the TESS pixels, approximately 21\arcsec\ on a side, leave it susceptible to photometric contamination from nearby stars, including additional wide stellar companions. We searched for nearby sources to TIC~280206394 with SOAR speckle imaging \citep{tokovinin:2018} on 18 May 2019 UT, observing through a similar visible bandpass as TESS. More details of these observations are available in \citet{ziegler:2019}. We detected no nearby sources within 3\arcsec\ of TIC~280206394. The 5$\sigma$ detection sensitivity and the speckle auto-correlation function from the SOAR observation are plotted in Figure~\ref{fig:soar}.

\medskip

The radial velocity variations measured on \stname\ phase with the transit signal. This fact, combined with the lack of nearby companions, the lack of correlation of the bisector span measurements with orbital phase and the tests carried out as part of the SPOC data validation report show that the transit is not caused by a blended stellar eclipsing binary.

\section{Analysis} \label{sec:ana}

\subsection{Stellar parameters}
\label{ssec:stellarpars}
In order to characterize the star, we follow the same procedure presented in \citet{brahm:2019}.
First, we compute the stellar atmospheric parameters
using the co-added FEROS spectra through the \zaspe\
code \citep{brahm:2016:zaspe}.
ZASPE estimates \teff, \logg, \feh, and \vsini, by comparing an observed spectrum
with a grid of synthetic models generated with the ATLAS9 atmospheres \citep{atlas9}.

Then, we estimate the physical parameters of the host star using the publicly available broadband photometry of GAIA (G, B$_P$, R$_P$) and 2MASS (J, H, K$_S$), which is compared to the synthetic magnitudes provided by the PARSEC stellar evolutionary models by using the distance to the star from the Gaia DR2 parallax. For a given stellar mass, age and metallicity, the PARSEC models
can deliver a set of synthetic absolute magnitudes and other stellar properties (e.g. stellar luminosity, effective temperature, stellar radius).

We determine the posterior distributions for \mstar, Age, and A$_V$, via an MCMC code using the \texttt{emcee} package \citep{emcee:2013}, where we fix the metallicity of the PARSEC models to that obtained with \zaspe, and we
apply the \citet{cardelli:89} extinction laws to the synthetic magnitudes.

This procedure provides a more precise estimation of \logg\  than the one obtained from the spectroscopic analysis. For this reason we iterate the procedure where we fix the \logg\ value when running \zaspe\ to the value obtained from the PARSEC models. The resulting values are $\teff = \stteff$ K, $\feh = \stfeh$ dex, $\logg = \stlogg$, $\vsini = \stvsini$ \kms, $A_V = \av$ mag, age $=\stage$ Gyr, $L_{\star}=\stlum$ \lsun\  $\mstar = \stmstan$ \msun\ and $\rstar = \strstan$ \rsun. The values and uncertainties of \mstar\ and \rstar\ are used to define priors for them in the global analysis described in the next section.

\begin{deluxetable*}{lrc}[b!]
\tablecaption{Stellar properties of \stname\  \label{tab:stprops}}
\tablecolumns{3}
\tablewidth{0pt}
\tablehead{
\colhead{Parameter} &
\colhead{Value} &
\colhead{Reference} \\
}
\startdata
Names \dotfill   &    \stnameTIC  & TIC  \\
 & 2MASS J09362869-5027478 & 2MASS  \\
 & TYC 8176-02431-1 & TYCHO  \\
 & WISE J093628.65-502747.3 & WISE  \\
RA \dotfill (J2000) &  15h32m17.84s &  \\
DEC \dotfill (J2000) & -22d21m29.74s &   \\
$\mu_\alpha$ \hfill (mas yr$^{-1}$) & -24.82 $\pm$ 0.05 & Gaia\\
$\mu_\delta$ \dotfill (mas yr$^{-1}$) & 42.42 $\pm$ 0.05 & Gaia\\
$\pi$ \dotfill (mas)& $7.02 \pm 0.03$  & Gaia \\
\hline
TESS \dotfill (mag)  & 9.24 $\pm$ 0.018 & TIC\\
G  \dotfill  (mag) & 9.661 $\pm$ 0.020 & Gaia\\
B$_P$ \dotfill  (mag) & 9.968 $\pm$ 0.005 & Gaia\\
R$_P$ \dotfill  (mag) & 9.229 $\pm$ 0.003 & Gaia\\
J  \dotfill (mag) & 8.722 $\pm$ 0.020 & 2MASS\\
H  \dotfill (mag) & 8.470 $\pm$ 0.038 & 2MASS\\
K$_s$  \dotfill (mag) & 8.429 $\pm$ 0.023 & 2MASS\\
\hline
\teff  \dotfill (K) & \stteff & \texttt{zaspe}\\
\logg \dotfill (dex) & \stlogg & \texttt{zaspe}\\
\feh \dotfill (dex) & \stfeh & \texttt{zaspe}\\
\vsini \dotfill (km s$^{-1}$) & \stvsini & \texttt{zaspe}\\
\mstar \dotfill (\msun) & \stm & this work\\
\rstar \dotfill (\rsun) & \str & this work \\
Age \dotfill (Gyr) & \stage & this work\\
\rhostar \dotfill (g cm$^{-3}$) & \strho & this work\\
\enddata
\end{deluxetable*}

\subsection{Global modeling}
\label{sec:glob}

We performed joint modelling of the radial velocity and photometric data using the
\textsf{exoplanet} toolkit \citep{exoplanet:exoplanet}. The radial velocities used are given in Table~\ref{tab:rvs} and the photometric data are given in Table~\ref{tab:phot}.
We denote the TESS photometric time series
by $\{y_T(t_i)\}_{i=1}^{n_T}$, the SSMKO one by $\{y_S(t_i)\}_{i=1}^{n_S}$, and the radial velocity measurements (with their mean values removed) by $\{y_{\rm F}(t_i)\}_{i=1}^{n_F}$,
$\{y_{\rm C}(t_i)\}_{i=1}^{n_C}$, $\{y_{\rm \chi}(t_i)\}_{i=1}^{n_\chi}$, $\{y_{\rm N}(t_i)\}_{i=1}^{n_N}$ and $\{y_{\rm M(i)}(t_i)\}_{i=1}^{n_N}$ for FEROS, Coralie, CHIRON,  NRES, and Minerva Australis respectively. In the case of Minerva, $M(i)$ is a function that returns the telescope used at observation $t_i$ (recall the Minerva observations include three different telescopes). The observational uncertainties are denoted by $\sigma_*(t_i)$, where $*$ can take the value of any of the instrument labels. As shown in \S~\ref{ssec:tess}, the TESS light curve shows no evidence of additional structure beyond white noise. The TESS photometric time series is therefore modeled as

\begin{align}
    y_T(t_i) &= \mathcal{T}(t_i; \mathbf{p}) + N(0,\sigma^2_{T,i}),
\end{align}

\noindent where $N(0,\sigma^2)$ denotes a Normal distribution of mean 0 and variance $\sigma^2$, $\mathcal{T}(t_i; \mathbf{p})$ is the transit model, and $\mathbf{p}$ is the vector of
model parameters. Explicitly,
\begin{align}
\mathbf{p} &= (R_p,b,P, t_0, e, \omega, R_*, M_*, u_1, u_2),
\end{align}
\noindent where $R_p$ is the planetary radius, $b$ is the impact parameter, $P$ is the period, $t_0$ the reference time of mid-transit,  $e$ the eccentricity, $\omega$ the angle of periastron, $R_*$ and $M_*$ the stellar radius and mass, and $u_1$ and $u_2$ the limb-darkening law coefficients, which we describe using a quadratic law.

The SSMKO photometric time series is modeled as
\begin{align}
    y_S(t_i) &= \mathcal{T}(t_i; \mathbf{p_{u}}) + N(0,\sigma^2_{S,i} + \eta_S^2) \nonumber \\
             &+ b_0 + b_1(t_i-t_0) \nonumber \\
             &\equiv \phi_S(t_i) + N(0,\sigma^2_{S,i}),
\end{align}
\noindent where the $\{b_i\}$ coefficients accounts for up to a linear systematic trend in the photometry, $\sigma^2_{S,i}$ are the reported photometric uncertainties, and $\eta_S$ is an additional photometric variance parameter. The parameter vector is the same as that for TESS, but the limb darkening coefficients are fixed to the values $(u_1,u_2)=(0.2489, 0.305)$. These values were calculated using the ATLAS atmospheric models and the Sloan i' band using the limb darkening coefficient calculator \citep{espinoza:2015}, and in particular using the methodology of sampling the limb darkening profile in 100 points as described in \citet{espinoza:2015}. We chose to fix the limb darkening coefficients given that the SSMKO light curve covers only the egress.
The radial velocity times series are modeled as

\begin{align}
    y_*(t_i) &= \mathcal{O}(t_i; \mathbf{q}) + N(0,\sigma^2_{*,i} + \eta^2_{*}) + \gamma_* \nonumber\\
          & a_0 + a_1(t-t_0)  \nonumber\\
          &\equiv r_*(t_i) + N(0,\sigma^2_{*,i} + \eta^2_{*})
\end{align}
where $\mathcal{O}$ represents the Keplerian radial velocity curve and the parameter vector of
the model, a subset of $\mathbf{p},$ is $\mathbf{q} = (P, t_0, e, \omega, M_*)$. The wildcard $*$ can take the values ${F,C,\chi,N,M}$ for FEROS, Coralie, CHIRON, NRES and Minerva respectively, and $\eta^2_*$ is a white noise ``jitter" terms to account for additional variance not accounted for in the observational uncertainties. The parameters ${a_i}$ account for up to a linear systematic trend in the radial velocities. We set priors for $\eta^2_*$ by first running a model without jitter terms, and determining for each instrument how much extra variance was present around the posterior model over that predicted by the observational uncertainties. We note that for NRES we found no need for a jitter term and thus we set $\eta_N \equiv 0$.
The log-likelihood $l$ is given by

\begin{align}
    -2 l(\mathbf{p}) &= \medmath{\sum_{i=1}^{n_T}\sigma_{T,i}^{-2}(y_{T,i}-\mathcal{T}_{T,i})^2} \nonumber\\
                    &\medmath{+ \sum_{i=1}^{n_S}(\sigma_{S,i}^{2} + \eta_S^2)^{-2}(y_{S,i}-\phi_{S,i})^2 + \ln(\sigma_{S,i}^{2} + \eta_S^2)}\nonumber\\
                    &\medmath{+ \sum_{i=1}^{n_F}(\sigma_{F,i}^2+\eta^2_F)^{-2}(y_{F,i}-r_{F,i})^2 +\ln(\sigma_{F,i}^2+\eta^2_F)}\nonumber\\
                    &\medmath{+ \sum_{i=1}^{n_C}(\sigma_{C,i}^2+\eta_C^2)^{-2}(y_{C,i}-r_{C,i})^2 +\ln(\sigma_{C,i}^2+\eta_C^2)}\nonumber\\
                    &\medmath{+ \sum_{i=1}^{n_\chi}(\sigma_{\chi,i}^2 + \eta^2_\chi)^{-2}(y_{\chi,i}-r_{\chi,i})^2 +\ln(\sigma_{\chi,i}^2 + \eta^2_\chi)} \nonumber\\
                    &\medmath{+\sum_{i=0}^{n_M}(\sigma_{M(i),i}^2 + \eta^2_M)^{-2}(y_{M,i}-r_{M(i),i})^2
                    + \ln(\sigma_{M(i),i}^2 + \eta^2_M)} \nonumber\\
                    &\medmath{+ \sum_{i=1}^{n_N}\sigma_{N,i}^{-2}(y_{N,i}-r_{N,i})^2}.
\end{align}

Posteriors were sampled using an Monte Carlo Markov Chain algorithm, specifically the No U-Turn Sampler \citep[NUTS,][]{hoffman:2011} as implemented in the \textsf{PyMC3} package through \textsf{exoplanet}.
We sampled using 4 chains and 3000 draws, after a tuning run of 4500 draws where the step sizes are optimized. Convergence was verified using the Rubin-Gelman and Geweke statistics. The effective sample size for all parameters, as defined by \citep{gelman:2013:bda}, was $>4000$.
The priors are detailed in Table~\ref{tab:plprops}. Priors for $M_*$ and $R_*$ stem from the analysis described in \S~\ref{ssec:stellarpars}. The priors on $P$, $T_0$, $\ln(\mpl/m\mearth)$, $e$, $\{a_i\}_{i=0}^1$ were obtained from a fit to the radial velocities alone carried out with the \textsf{radvel} package \citep{fulton:2018}.

\begin{deluxetable}{lrc}[b!]
\tablecaption{Prior and posterior parameters of the global fit. Derived parameters which are deterministic functions of the parameters fitted for are presented in the bottom part of the table. For the priors, $N(\mu,\sigma^2)$ stands for a normal distribution with mean $\mu$ and variance $\sigma^2$, $N(\mu,\sigma^2; l, u)$ is a bounded normal distribution with lower and upper limits given by $l$ and $u$, respectively, and $U(a,b)$ stands for a uniform distribution between $a$ and $b$.\label{tab:plprops}}
\tablecolumns{3}
\tablenum{2}
\tablewidth{0pt}
\tablehead{
\colhead{Parameter} &
\colhead{Prior} &
\colhead{Value} \\
}
\startdata
P (days) & $N(11.27,0.7^2)$  & \per \\
T$_0$ (BJD)&  $N(2458547.4743,0.0012^2)$ &  \tzero \\
$\ln (\mpl/\mearth)$ & $N(\ln(500),0.3^2)$ & \logm \\
\rstar\ (\rsun)& $N(1.282, 0.03^2)$  & \str \\
\mstar\ (\msun) &  $N(1.17,0.06^2) $& \stm  \\
r1\tablenotemark{a} & $U(0,1)$ & \rone \\
r2\tablenotemark{a} & $U(0,1)$ & \rtwo \\
u$_1^{\rm TESS}$ & $U(0,1)$ & \uone \\
u$_2^{\rm TESS}$ & $U(0,1)$& \utwo \\
$e$ & $N_b(0.4,0.1; 0,1)$ & \ecc \\
$\omega$ (rad) & $U(-\pi,\pi)$ & \omegapost \\
$\gamma_{\rm FEROS}$  (m s$^{-1}$)& $N(-40,20^2)$& \gammaferos \\
$\gamma_{\rm Coralie}$ (m s$^{-1}$)& $N(-20,20^2)$ & \gammacoralie \\
$\gamma_{\rm CHIRON}$ (m s$^{-1}$)& $N(-20,20^2)$ & \gammachiron \\
$\gamma_{\rm NRES}$ (m s$^{-1}$)& $N(-20,20^2)$ & \gammanres \\
$\gamma_{\rm Minerva\_T3}$ (m s$^{-1}$)& $N(0,20^2)$ & \gammaminervathree \\
$\gamma_{\rm Minerva\_T4}$ (m s$^{-1}$)& $N(-20,20^2)$ & \gammaminervafour \\
$\gamma_{\rm Minerva\_T5}$ (m s$^{-1}$)& $N(0,20^2)$ & \gammaminervafive \\
$\eta_{\rm FEROS}$ (m s$^{-1}$)& $N_b(13,5^2; 0,\infty)$ & \etaferos \\
$\eta_{\rm Coralie}$ (m s$^{-1}$)& $N_b(22,10^2; 0, \infty)$ & \etacoralie \\
$\eta_{\rm CHIRON}$ (m s$^{-1}$)& $N_b(45,15^2; 0, \infty)$ & \etachiron \\
$\eta_{\rm Minerva}$ (m s$^{-1}$)& $N_b(30,12^2; 0, \infty)$ & \etaminerva \\
$\eta_{\rm S}$ & $N_b(0.001,0.0003^2; 0, \infty)$ & \etassmko \\
$a_0$ (m s$^{-1}$) & $N(0,20^2)$ & \azero\\
$a_1$ (m s$^{-1}$ d$^{-1}$) & $N(0.7,0.5^2)$ & \aone\\
$b_0$  & $N(0,0.1^2)$ & \bzero\\
$b_1$ (d$^{-1}$)& $N(0,0.01^2)$ & \bone\\
\hline
$b$ &  & \bb \\
$i$ (deg) &  & \inc \\
\rpl/\rstar  &  & \rr \\
\rpl (\rjup)& & \rptessrjup\\
%\mpl (\mearth) & & \mptessEarth\\
\mpl (\mjup) & & \mptess\\
$a$ (AU)    & & \sma \\
\teq (K) \tablenotemark{b}   & & \teqv \\
\enddata
\tablenotetext{a}{These parameters correspond to the parametrization presented in \citet{exoplanet:espinoza18} for sampling physically possible combinations of $b$ and
p = \rpl/\rstar. We used an upper and lower allowed value for $p$ of $p_l=0.075$ and $p_u=0.125$, respectively.}
\tablenotetext{b}{Time-averaged equilibrium temperature computed according to equation~16 of \citet{mendez:2017}}
\end{deluxetable}

\begin{figure*}
\plotone{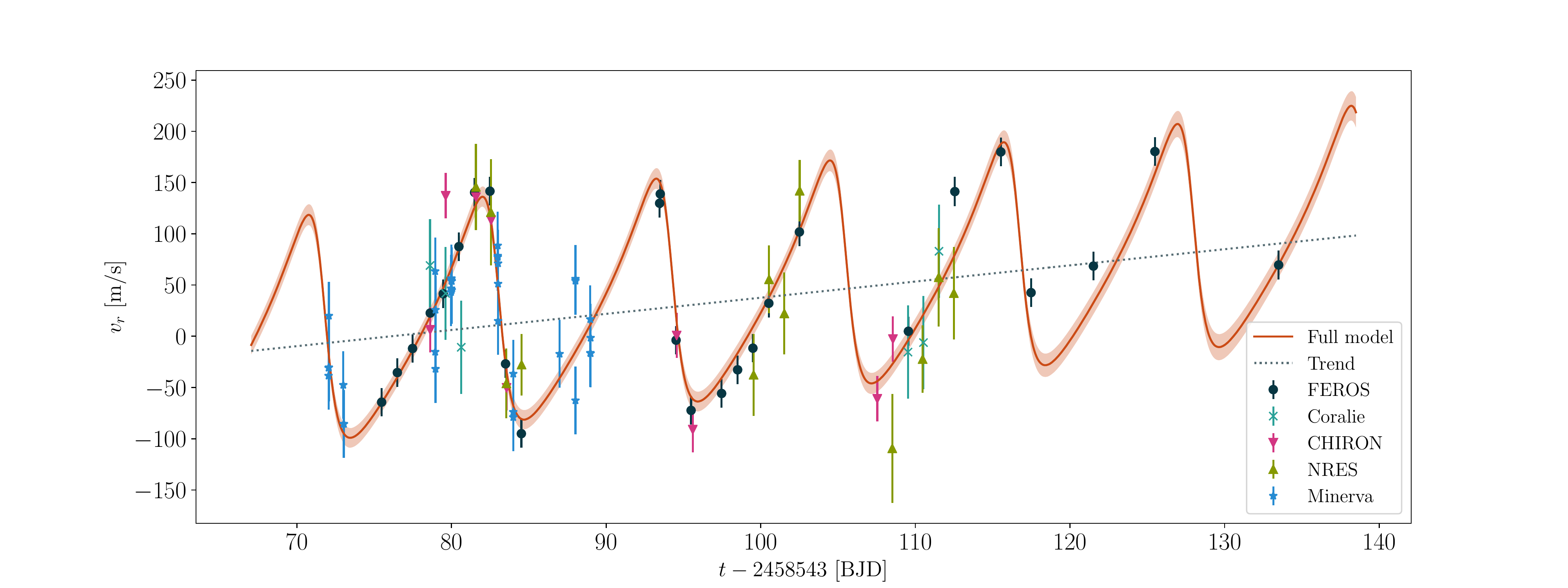}
\caption{Radial velocities as a function of time for the five spectroscopic instruments used in this work (FEROS, Coralie, CHIRON, NRES, Minerva). Note that we use a single symbol for Minerva but the observations were made with three different telescopes in the array. The error bars include the posterior values for the jitter terms.
}
\label{fig:rvs_time}
\end{figure*}

\begin{figure}
\plotone{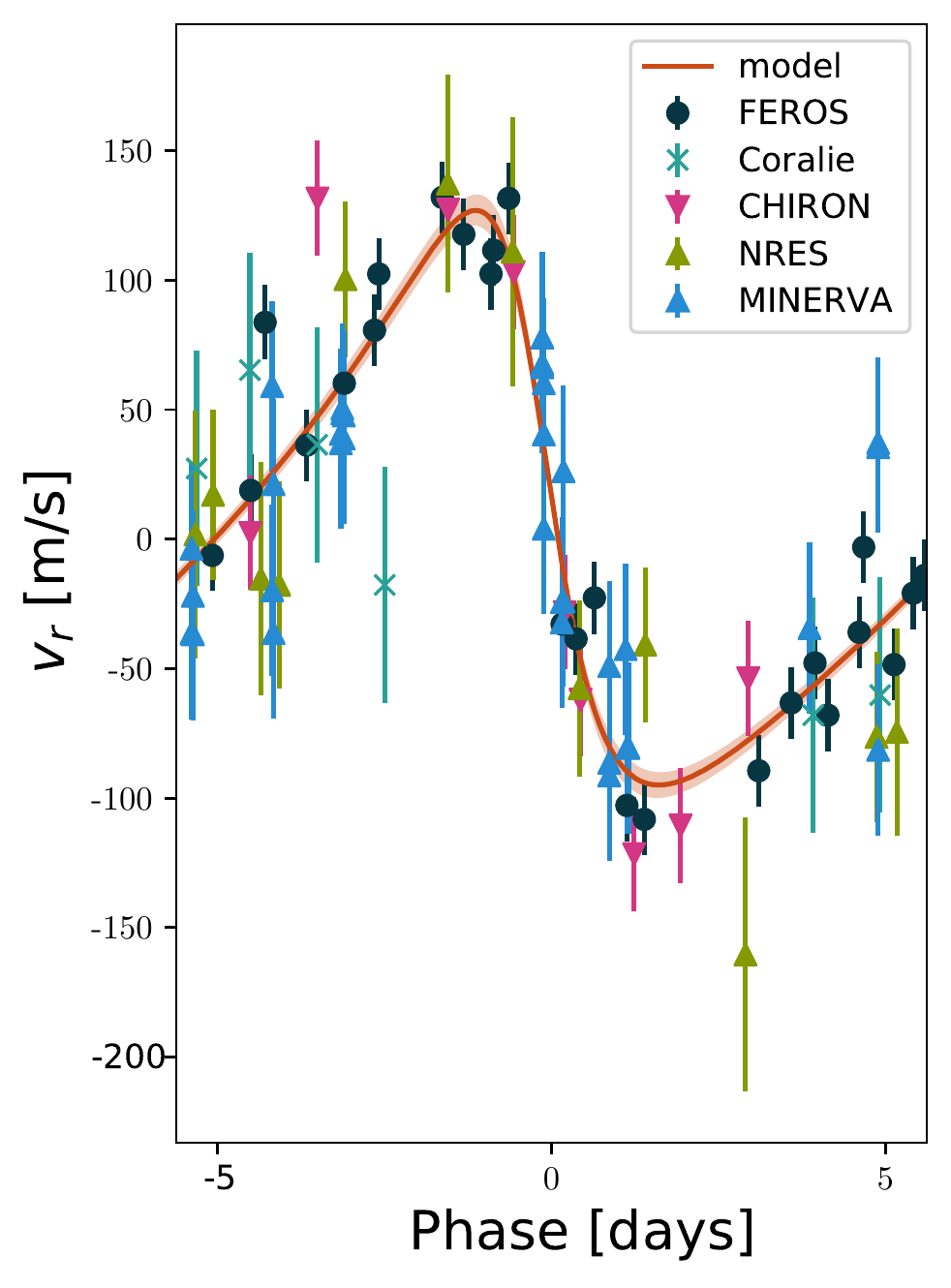}
\caption{Radial velocities as a function of orbital phase for the five spectroscopic instruments used in this work (FEROS, Coralie, CHIRON, NRES, Minerva). Note that we use a single symbol for Minerva but the observations were made with three different telescopes in the array. The error bars include the posterior values for the jitter terms.
}
\label{fig:rvs_phased}
\end{figure}

The posterior model for the radial velocities is shown in Figure~\ref{fig:rvs_time} as a function of time and in Figure~\ref{fig:rvs_phased} against orbital phase with the quadratic term removed. The posterior model for the photometric observations is shown in Figure~\ref{fig:lcs_phased}. Table~\ref{tab:plprops} lists all the priors assumed and the posterior values for the stellar and planetary properties.
A fully independent analysis of the data with the \textsf{juliet} package \citep{espinoza:juliet} using different priors and treatment of photometric and radial velocity trends results in planetary parameters consistent with the ones presented in Table~\ref{tab:plprops}.
It is noteworthy that besides the Keplerian orbit, there is significant statistical evidence for a long term trend in the radial velocities which could be caused by an outer companion. If described by a linear trend the slope is estimated to be $a_1=\aone$ m s$^{-1}$ d$^{-1}$.

\vfill

\section{Discussion} \label{sec:dis}

\begin{figure*}
\plotone{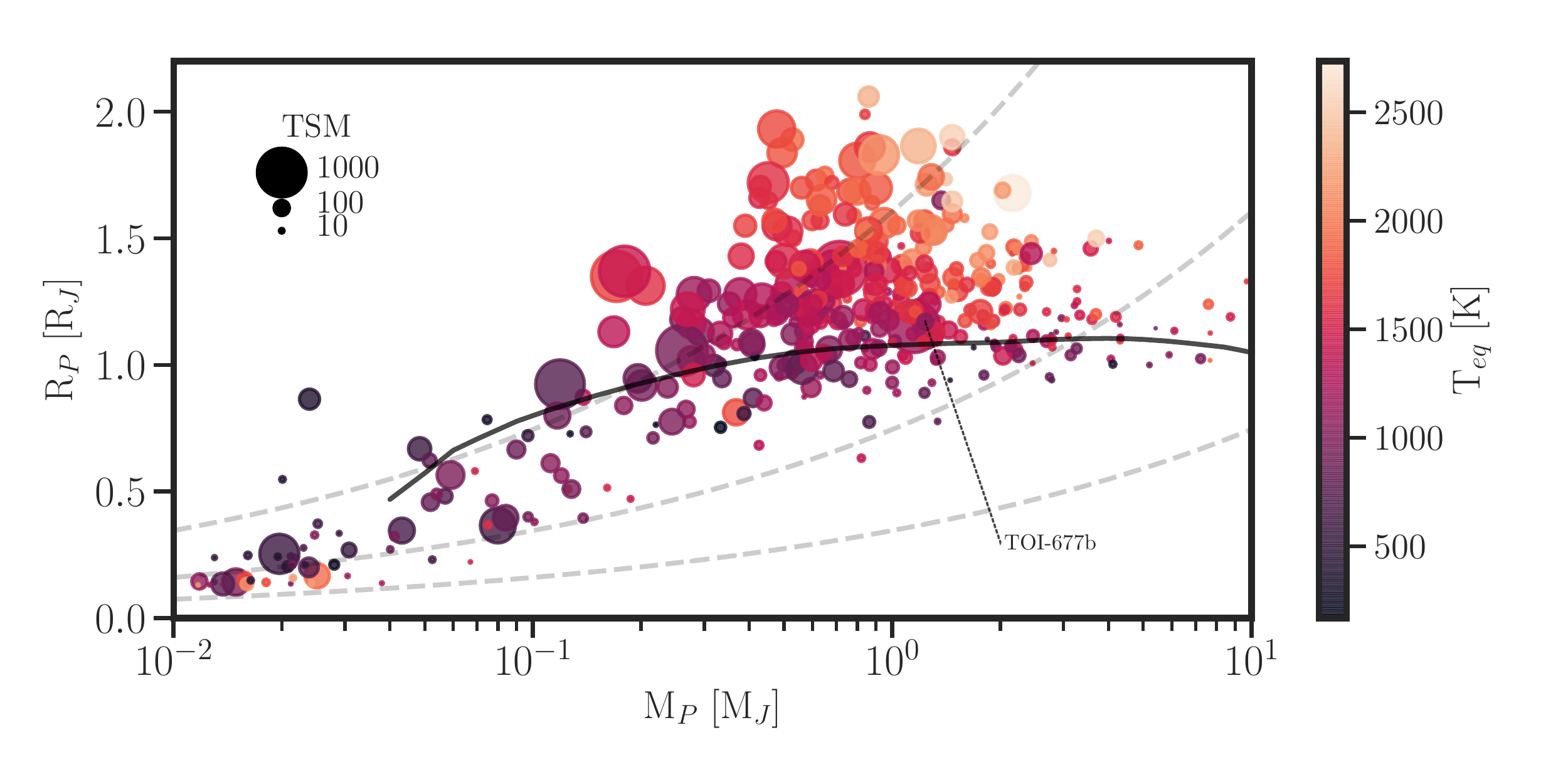}
\caption{Mass -- Radius diagram for the population of well characterized transiting planets \citep{southworth:2011:tepcat}. The point corresponding to \plname\ is indicated with a dashed line. The color represents the equilibrium temperature of the planet, while the size scales down with the transmission spectroscopy metric as defined by \citet{tsm}. The dashed gray lines correspond to isodensity curves for 0.3, 3 and 30 g cm$^{-3}$, respectively. The solid line corresponds to the predicted radius using the models of \citet{fortney:2007} for a planet with a 10~\mearth\ central core.\label{fig:mr}}
\end{figure*}

We put \plname\ in the context of the population of known, well characterized\footnote{We use the catalog of well characterized planets of \citep{southworth:2011:tepcat}. We restrict the sample to systems whose fractional error on their planetary masses and radii are $<25\%$.} transiting exoplanets in Figure~\ref{fig:mr}, where we show a scatter plot of planetary mass versus planetary radius, coding with color the equilibrium temperature. The incident flux for \plname\ is $\approx 3\times 10^8$ erg s$^{-1}$ cm$^{-2}$, very close to the value of $\approx 2\times 10^8$ erg s$^{-1}$ cm$^{-2}$ below which it has been shown that the effects of irradiation on the planetary radius are negligible \citep[e.g.][]{demory:2011}. The radius of \plname\ is in line with what is expected for a gas giant with a core of $M_C=10\mearth$ according to the standard models of \citet{fortney:2007}. This underscores the value of warm giants, whose structure can be modeled without the complications of an incident flux resulting in radius inflation \citep{kovacs:2010,demory:2011}.
Figure~\ref{fig:mr} also shows that \plname, having a transmission spectroscopy metric \citep[TSM,][]{tsm} of $\approx$100, is not a particularly
well suited target of transmission spectroscopy studies, if compared with the rest of the population of close-in giant planets.

\begin{figure*}
\plotone{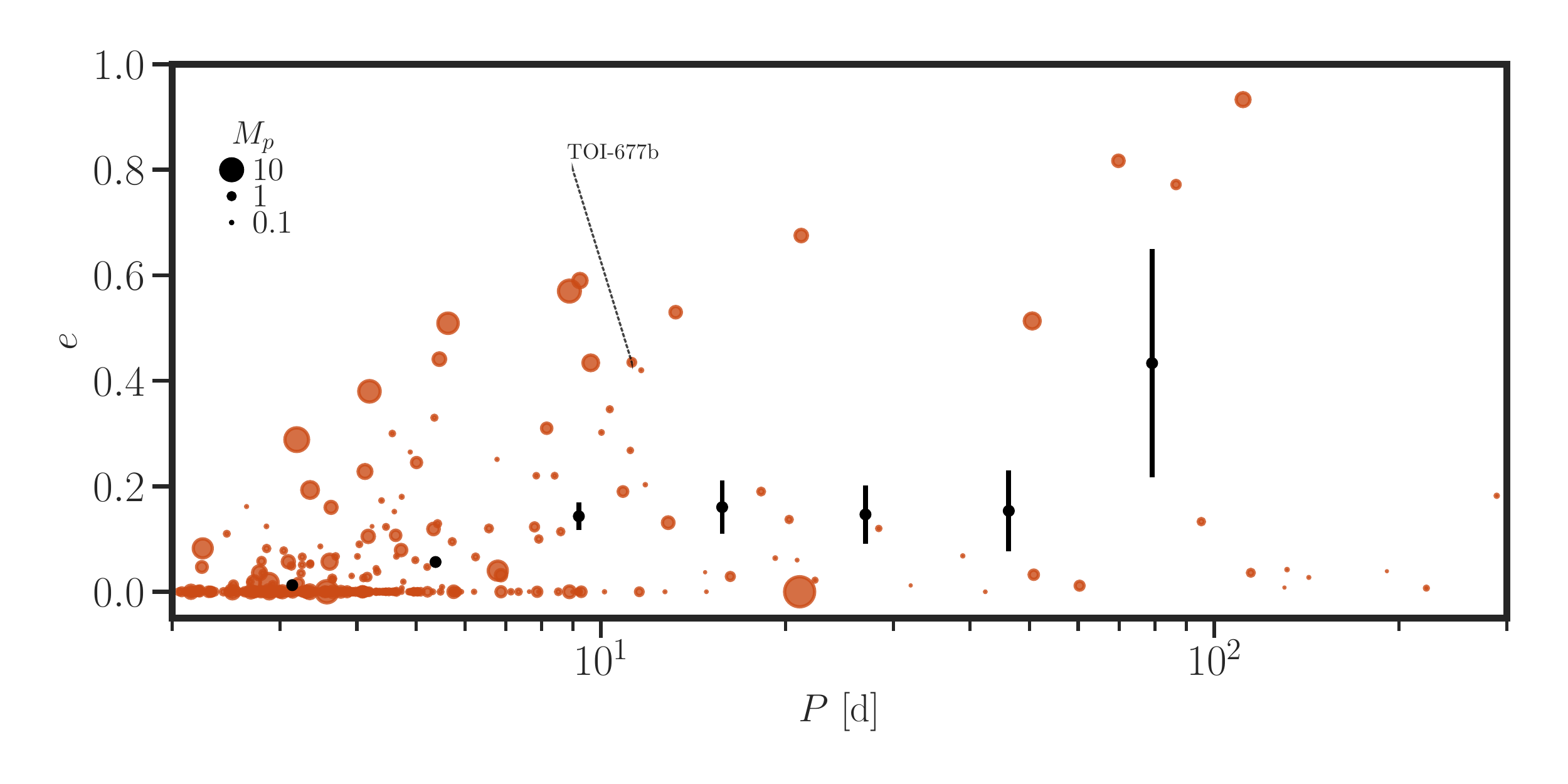}
\caption{Period -- Eccentricity diagram for the population of well characterized transiting planets. The point corresponding to \plname\ is indicated with a dashed line. The size scales with the mass of the planet. The black points with error bars are the average eccentricities of the sample in bins that are equally spaced in $\log(P)$ for $P<100$ d. \label{fig:eccP}}
\end{figure*}

In Figure~\ref{fig:eccP} we plot the same population of well-characterized planets in the period--eccentricity plane, coding the planetary mass with the symbol size. It is apparent that \plname\ lies in a part of this plane that is still sparsely populated. The eccentricity of exoplanets is very low for close-in systems, and starts to grow for periods $P\gtrsim 4$ d. With an eccentricity of $e=\ecc$, \plname\ lies in the upper range of eccentricity values for planets  with similar periods in the currently known sample. Besides the significant eccentricity of the orbit of \plname\, the presence of a long term trend
in the radial velocities is interesting in the context of migration mechanisms of giant planets. Warm jupiters can be formed via secular gravitational interactions with an outer planet followed by tidal interactions with the star in the high eccentricity stage of the secular cycle \citep[e.g. ][]{kozai:62}. In this context, \citet{dong:2014} predicts that
in order to  overcome  the  precession  caused  by  general  relativity, the warm jupiters produced via this mechanism should have outer planets at relatively short orbital distances that can be detected with a radial velocity monitoring.
At the moment we cannot provide meaningful constraints on a potential outer companion. We will continue to monitor the system with radial velocities to determine the exact nature of the long term radial velocity we uncovered.

The determination of the orbital obliquity of transiting planets through the Rossiter-McLaughlin (R-M) effect, particularly for planets with orbital periods longer than $\approx$10 days, provides a powerful tool to constrain migration theories \citep{petrovich:2016}. With a sizable \vsini\ of \stvsini\ \kms\ and a bright magnitude of $V=9.8$ mag, \plname\ is a prime target to perform a measurement of the projected angle between the stellar and orbital angular momenta. Specifically, the expected semi-amplitude of the R-M signal for \plname\ in the case of an aligned orbit is of
K$_{R-M} = 70 \pm 10$ ms$^{-1}$.
While still based on a very limited population, the current obliquity distribution of transiting planets with similar periods as \plname\ seems to follow a similar behaviour to that of the eccentricity distribution, with a large spread in their values. Current discoveries include aligned systems like WASP-84b \citep{wasp84} and HAT-P-17b \citep{fulton:2013}, mildly misaligned systems \citep[WASP-117b,][]{wasp117}, and also others that are even retrograde \citep[WASP-8b,][]{queloz:2010}. The measurement of the obliquity of \plname\ will increment this small sample and help in further understanding how close-in giant planets form.

\acknowledgements

A.J.\ acknowledges support from FONDECYT project 1171208 and by the Ministry for the Economy, Development, and Tourism's Programa Iniciativa Cient\'{i}fica Milenio through grant IC\,120009, awarded to the Millennium Institute of Astrophysics (MAS).
R.B.\ acknowledges support from FONDECYT Post-doctoral Fellowship Project 3180246, and from the Millennium Institute of Astrophysics (MAS).
M.B.\ acknowledges CONICYT-GEMINI grant 32180014.
T.D.\ acknowledges support from MIT’s Kavli Institute as a Kavli postdoctoral fellow.
Resources supporting this work were provided by the NASA High-End Computing (HEC) Program through the NASA Advanced Supercomputing (NAS) Division at Ames Research Center for the production of the SPOC data products.
This work has made use of data from the European Space Agency (ESA) mission Gaia (\url{https:
//www.cosmos.esa.int/gaia}), processed by the Gaia Data
Processing and Analysis Consortium (DPAC, \url{ https://www.cosmos.esa.int/web/gaia/dpac/consortium}). Funding for the DPAC has been provided by national ins
titutions, in particular
the institutions participating in the Gaia Multilateral
Agreement.
This research made use of \textsf{exoplanet} \citep{exoplanet:exoplanet} and its
dependencies \citep{exoplanet:astropy13, exoplanet:astropy18,
exoplanet:espinoza18, exoplanet:exoplanet, exoplanet:kipping13,
exoplanet:luger18, exoplanet:pymc3, exoplanet:theano}.
MINERVA-Australis is supported by Australian Research Council LIEF Grant LE160100001, Discovery Grant DP180100972, Mount Cuba Astronomical Foundation, and institutional partners University of Southern Queensland, UNSW Australia, MIT, Nanjing University, George Mason University, University of Louisville, University of California Riverside, University of Florida, and The University of Texas at Austin.
We respectfully acknowledge the traditional custodians of all lands throughout Australia, and recognise their continued cultural and spiritual connection to the land, waterways, cosmos, and community. We pay our deepest respects to all Elders, ancestors and descendants of the Giabal, Jarowair, and Kambuwal nations, upon whose lands the Minerva-Australis facility at Mt.\ Kent is situated.

\vspace{5mm}
\facilities{TESS, FEROS/MPG2.2m, Coralie/Euler1.3m, NRES/LCOGT1m, Minerva Australis, Shared Skies Telescope/Mount Kent Observatory, CTIO 1.5m/CHIRON}

\software{
          CERES \citep{brahm:2017:ceres,jordan:2014},
          ZASPE \citep{brahm:2016:zaspe,brahm:2015},
          \textsf{radvel} \citep{fulton:2018},
          \textsf{exoplanet} \citep{exoplanet:exoplanet},
          \textsf{juliet} \citep{espinoza:juliet}
          }

%\clearpage
%\bibliographystyle{apj}
\bibliography{tessbib}

\appendix

\begin{deluxetable*}{lrrl}
\tablewidth{0pc}
\tablecaption{Relative photometry for \stname.\tablenotemark{a}\label{tab:phot}}
\tablehead{
\colhead{BJD} & \colhead{f} & \colhead{$\sigma_{\rm f}$} &  \colhead{Instrument} \\
\colhead{\hbox{(2,400,000$+$)}} & \colhead{ppt} & \colhead{ppt}  & \colhead{}
}
\startdata
 58547.001330 & -0.199 &  0.789 & TESS \\
 58547.002719 &  1.058 &  0.790 & TESS \\
 58547.004108 &  0.339 &  0.789 & TESS \\
 58547.005496 & -1.082 &  0.790 & TESS \\
 58547.006885 &  0.377 &  0.790 & TESS \\
 58547.008274 &  1.051 &  0.790 & TESS \\
 58547.009663 &  0.314 &  0.789 & TESS \\
 58547.011052 &  1.058 &  0.790 & TESS \\
 58547.012441 & -0.483 &  0.790 & TESS \\
 58547.013830 &  0.617 &  0.789 & TESS \\
 58547.015219 & -0.675 &  0.790 & TESS \\
 58547.016608 & -0.039 &  0.789 & TESS \\
 58547.017997 &  0.166 &  0.790 & TESS \\
 58547.019386 &  0.393 &  0.788 & TESS \\
 58547.020774 & -0.916 &  0.790 & TESS \\
 58547.022163 & -1.167 &  0.789 & TESS \\
 58547.023552 &  2.311 &  0.790 & TESS \\
 58547.024941 &  1.251 &  0.790 & TESS \\
 58547.026330 &  0.043 &  0.789 & TESS \\
 58547.027719 & -0.603 &  0.790 & TESS
\enddata
\tablenotetext{a}{ Table~\ref{tab:phot} is published in its entirety in
machine readable format. A portion is shown here for guidance regarding its form and content.
}
\end{deluxetable*}

\startlongtable
\begin{deluxetable*}{lrrrrl}
\tablewidth{0pc}
\tablecaption{Radial velocities and bisector spans for \stname.\tablenotemark{a}}
\tablehead{
\colhead{BJD} & \colhead{RV\tablenotemark{b}} & \colhead{$\sigma_{\rm RV}$} & \colhead{BIS} & \colhead{$\sigma_{\rm BIS}$} & \colhead{Instrument} \\
\colhead{\hbox{(2,400,000$+$)}} & \colhead{(m s$^{-1}$)} & \colhead{(m s$^{-1}$)} & \colhead{(m s$^{-1}$)} & \colhead{(m s$^{-1}$)} & \colhead{}
}
\startdata
 58615.051551 & -26.62 & 5.4 & \ldots & \ldots & Minerva\_T3 \\
 58615.051551 & -70.76 & 5.3 & \ldots & \ldots & Minerva\_T4 \\
 58615.072962 & -26.25 & 5.4 & \ldots & \ldots & Minerva\_T3 \\
 58615.072962 & -12.28 & 5.3 & \ldots & \ldots & Minerva\_T4 \\
 58616.005135 & -43.36 & 5.4 & \ldots & \ldots & Minerva\_T3 \\
 58616.026546 & -81.33 & 5.4 & \ldots & \ldots & Minerva\_T3 \\
 58616.047945 & -81.42 & 5.4 & \ldots & \ldots & Minerva\_T3 \\
 58618.484021 & -107.13 & 12.0 &  34 &   9 & FEROS \\
 58619.499881 & -78.23 & 9.6 &  47 &   8 & FEROS \\
 58620.482581 & -54.73 & 11.3 &  14 &   9 & FEROS \\
 58621.616231 & 42.13 & 23.5 & -51 &  20 & Coralie \\
 58621.623201 & -14.62 & 13.3 & \ldots & \ldots & CHIRON \\
 58621.628071 & -20.13 & 13.3 &  23 &  10 & FEROS \\
 58621.948677 & -47.48 & 5.2 & \ldots & \ldots & Minerva\_T4 \\
 58621.948677 & 67.59 & 5.4 & \ldots & \ldots & Minerva\_T3 \\
 58621.970088 & -6.47 & 5.4 & \ldots & \ldots & Minerva\_T4 \\
 58621.970088 & -27.69 & 5.4 & \ldots & \ldots & Minerva\_T3 \\
 58622.467171 & -1.33 & 9.8 &  35 &   8 & FEROS \\
 58622.623371 & 14.93 & 23.1 &  86 &  20 & Coralie \\
 58622.626601 & 116.18 & 24.5 & \ldots & \ldots & CHIRON \\
\enddata
\tablenotetext{a}{ Table~\ref{tab:rvs} is published in its entirety in
machine readable format. A portion is shown here for guidance regarding its form and content.
}
\tablenotetext{b}{For convenience, the mean has been subtracted from the originally measured radial velocities for each instrument, and the instrument-dependent radial velocity zeropoints reported in Table~\ref{tab:plprops} are with respect to these mean-subtracted values. The mean values $m$ which should be added to recover the original measurements are $m_{\rm FEROS}=37656.23$, $m_{\rm Coralie}= 37665.27$, $m_{\rm CHIRON}=20.1$, $m_{\rm NRES}=38228.49$ and $m_{\rm Minerva}=37844.01$}
\label{tab:rvs}
\end{deluxetable*}

%\allauthors

\end{document}